\documentclass[10pt,letterpaper,twocolumn,prl,showpacs]{revtex4-1}
\usepackage{graphicx,color,textcomp}
\usepackage{mathrsfs,amsmath}   
\usepackage{psfrag,overpic,pst-all}
\usepackage{amssymb}
\usepackage{epstopdf}

\begin{document}

\title{Observations of spatiotemporal instabilities in the strong-driving regime of an AC-driven nonlinear Schr\"odinger system}

\author{Miles Anderson$^1$}
\author{Fran\c{c}ois Leo$^{1}$}
\author{St\'ephane Coen$^{1}$}
\author{Miro Erkintalo$^{1}$}
\email{m.erkintalo@auckland.ac.nz}
\author{Stuart G. Murdoch$^1$}

\affiliation{$^1$The Dodd-Walls Centre for Photonic and Quantum Technologies, Department of Physics, The University of Auckland, Auckland 1142, New Zealand}

\begin{abstract}
{Localized dissipative structures (LDS) have been predicted to display a rich array of instabilities, yet systematic experimental studies have remained scarce. We have used a synchronously-driven optical fiber ring resonator to experimentally study LDS instabilities in the strong-driving regime of the AC-driven nonlinear Schr\"odinger equation (also known as the Lugiato-Lefever model). Through continuous variation of a single control parameter, we have observed a string of theoretically predicted instability modes, including irregular oscillations and chaotic collapses. Beyond a critical point, we observe behaviour reminiscent of a phase transition: LDSs trigger localized domains of spatiotemporal chaos that invade the surrounding homogeneous state. Our findings directly confirm a number of theoretical predictions, and they highlight that complex LDS instabilities can play a role in experimental systems.}
\end{abstract}

\maketitle

The AC-driven nonlinear Schr\"odinger equation (NLSE) is a prototypical model equation for pattern formation. It has relevance to many physical systems~\cite{barashenkov_existence_1996, barashenkov_existence_1999}, including RF-driven plasma~\cite{morales_ponderomotive-force_1977, nozaki_chaotic_1985}, long Josephson junctions~\cite{ustinov_solitons_1998}, and easy-axis ferromagnets in external magnetic fields~\cite{kosevich_magnetic_1990}. In particular, it provides for the canonical description of Kerr nonlinear optical cavities~\cite{lugiato_spatial_1987, haelterman_dissipative_1992}. In that context, the model is commonly referred to as the ``Lugiato-Lefever'' equation~\cite{lugiato_spatial_1987}, and it has been successfully applied to various distinct configurations, including spatially diffractive cavities~\cite{lugiato_spatial_1987, firth_cavity_2002, firth_theory_2001, ackemann_chapter_2009}, fiber ring resonators~\cite{haelterman_dissipative_1992, leo_temporal_2010, jang_ultraweak_2013, jang_temporal_2015, jang_writing_2015, leo_dynamics_2013}, and monolithic microresonators~\cite{matsko_mode-locked_2011, coen_modeling_2013, chembo_spatiotemporal_2013, herr_temporal_2014}.

The AC-driven NLSE has fixed-point solutions corresponding to localized dissipative structures (LDSs). These are wave packets that sit on top of a homogeneous background~\cite{firth_theory_2001, ackemann_chapter_2009}, with potential applications as coherent optical frequency combs~\cite{herr_temporal_2014, brasch_photonic_2016, yi_soliton_2015, erkintalo_coherence_2014}, or as bits in all-optical buffers~\cite{barland_cavity_2002, leo_temporal_2010, jang_temporal_2015}. Besides stable stationary LDSs, theoretical analyses have revealed several dynamic instabilities~\cite{nozaki_chaotic_1985, barashenkov_existence_1996, gomila_excitability_2005, leo_dynamics_2013, godey_stability_2014}. Periodically oscillating LDSs, akin to the ubiquitous ``oscillons''~\cite{umbanhowar_localized_1996, Lioubashevski_oscillons_1999,arbell_temporally_2000}, represent the simplest examples, yet more complex instability behaviours, including chaotic oscillations and transient collapses, have also been predicted~\cite{nozaki_chaotic_1985, leo_dynamics_2013}. Unstable LDSs of the AC-driven NLSE can even trigger a chain reaction of proliferation, giving rise to dynamics analogous to a phase transition: nucleation of a domain of spatiotemporal chaos that invades the surrounding meta-stable homogeneous state~\cite{leo_dynamics_2013}. The front dynamics~\cite{pomeau_front_1986} associated with the resulting expansion of a chaotic domain bears resemblance to the rise of turbulence in wall-bound shear flows~\cite{avila_onset_2011, barkley_rise_2015}, and to the onset of spatiotemporal chaos in systems ranging from excited granular layers~\cite{losert_propagating_1999} and prey-predator interactions~\cite{petrovskii_wave_2001} to liquid-crystal light valves~\cite{verschueren_spatiotemporal_2013, verschueren_chaotic_2014}.

Only stable~\cite{leo_temporal_2010, jang_ultraweak_2013, jang_temporal_2015, jang_writing_2015, herr_temporal_2014, brasch_photonic_2016, yi_soliton_2015, karpov_universal_2016} and periodically oscillating~\cite{leo_dynamics_2013} LDSs of the AC-driven NLSE have so far been studied in experiments. Observations of more complex behaviours have remained elusive, arguably due to the lack of appropriate experimental configurations. For example, whilst continuously-driven optical fibre ring resonators allow for the investigation of stable LDSs~\cite{leo_temporal_2010, jang_ultraweak_2013, jang_temporal_2015, leo_dynamics_2013}, they cannot easily access the strong driving regime where instabilities manifest themselves. Experiments involving microresonators~\cite{herr_temporal_2014, brasch_photonic_2016, yi_soliton_2015, karpov_universal_2016} routinely operate in that regime, but direct (e.g. time-resolved) observations are obstructed by the small system size. More fundamentally, lack of experiments has left open the question of whether the AC-driven NLSE remains a valid description for such systems in the strong-driving regime.

In this Letter, we experimentally investigate an optical fiber ring resonator that is synchronously driven with flat-top laser pulses. Through variation of a single control parameter, we observe a rich range of LDS instability dynamics that have been predicted to manifest themselves in the strong-driving regime of the AC-driven NLSE. This includes chaotically oscillating and collapsing LDSs, as well as transitions to expanding domains of spatiotemporal chaos. Our experimental findings directly confirm theoretically predicted LDS bifurcation characteristics, and they demonstrate that exotic instability behaviours can play a role in experimental systems.

\begin{figure}[t]
 \centering
  \includegraphics[width = 8.3cm, clip=true]{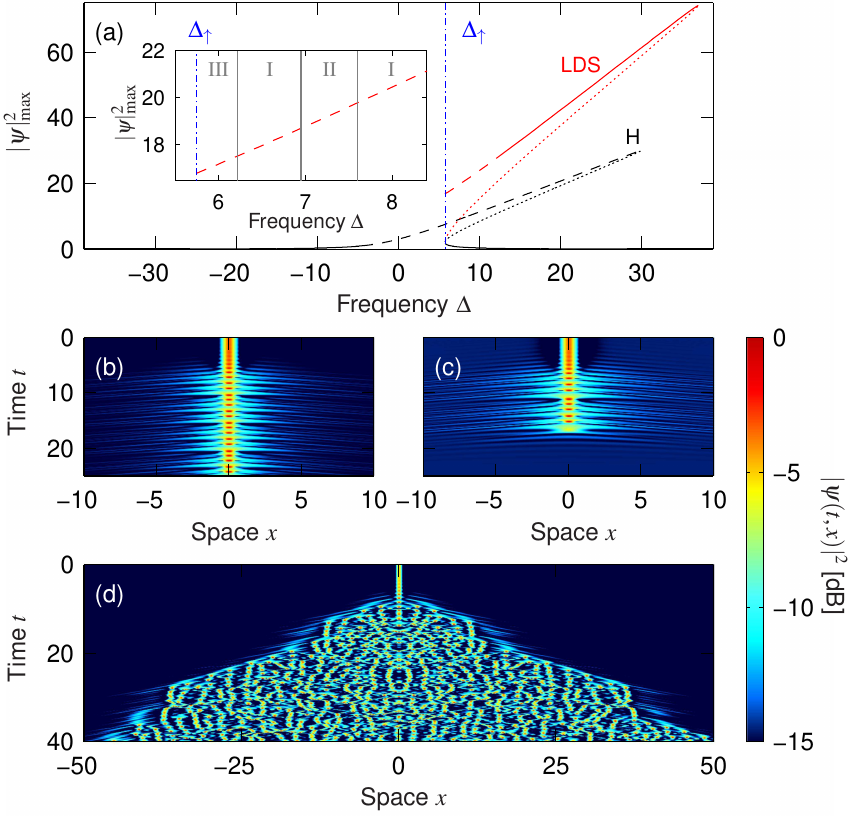}
 \caption{(a) Homogeneous (black curves) and LDS (red curves) solutions for $X = 30$. Solid curves: stable solutions; dotted curves: unstable solutions diverging to one of the other solutions; dashed lines: unstable solutions with more complex dynamics. Inset highlights the different dynamical LDS regimes: perpetually oscillating LDSs (I); oscillating LDSs that eventually decay to a homogeneous state (II); LDS that trigger spatiotemporal chaos (III). (b-d) Examples of numerically simulated dynamics for different regimes. (b) Regime I, $\Delta = 8$; (c) Regime II, $\Delta = 7.45$; (d) Regime III, $\Delta = 6$.}
 \label{Fig1}
\end{figure}

We first recall the theoretically predicted LDS instabilities~\cite{nozaki_chaotic_1985, barashenkov_existence_1996, leo_dynamics_2013} by considering the following dimensionless form of the AC-driven NLSE:
\begin{equation}
  \label{LLN}
  \frac{\partial \psi(t,x)}{\partial t} = \left[ -1 +i(|\psi|^2- \Delta)
  +i\frac{\partial^2}{\partial x^2}\right]\psi+S.
\end{equation}
Here $\psi(t,x)$ represents a field amplitude, whilst $S$ describes the strength of the homogeneous driver and $\Delta$ its frequency. Complex LDS instabilities have been predicted for $X  = |S|^2 \gtrsim 25$~\cite{leo_dynamics_2013}. In Fig.~\ref{Fig1} we illustrate generic bifurcation characteristics for a driving strength similar to the experiments that will follow ($X  = 30$). Here we show the peak amplitudes $|\psi|^2_\mathrm{max}$ of the co-existing homogenous ($H$, black curves) and LDS solutions (red curves) as a function of $\Delta$~\footnote{The solutions were obtained using standard techniques~\cite{leo_temporal_2010}.}. Considering first the homogeneous solutions, the intermediate branch (dotted line) is unconditionally unstable, while the top and bottom branches are stable against \emph{homogeneous} perturbations. Nevertheless, for $|\psi|^2>1$ \emph{inhomogeneous} perturbations can cause the top branch solutions (dashed line) to experience a Turing (or modulation) instability~\cite{lugiato_spatial_1987}, leading to stationary patterns or full spatiotemporal chaos consisting of fluctuating structures. The lower branch is stable for all $\Delta$, and corresponds to the coherent background on top of which the LDSs sit. Accordingly, the up-switching point $\Delta_\uparrow$ represents the lower boundary of LDS existence, as seen in Fig.~\ref{Fig1}(a). For LDS solutions, structures in the lower branch (dotted line) are metastable~\cite{leo_temporal_2010, jang_writing_2015}, and shall not be discussed further. On the upper branch, LDSs are stable (solid line) for large $\Delta$, but then become unstable (dashed line) through a Hopf bifurcation as $\Delta$ decreases.

Referring to the inset in Fig.~\ref{Fig1}(a), LDSs in the unstable regime first exhibit oscillatory breathing behaviour close to the Hopf threshold [region I, Fig.~\ref{Fig1}(b)]. The oscillations are initially periodic, but turn chaotic as $\Delta$ is reduced. For sufficiently small $\Delta$, behaviour not dissimilar to excitability~\cite{gomila_excitability_2005} emerges [region II, Fig.~\ref{Fig1}(c)]: the LDSs undergo a large excursion in phase space (chaotic oscillations), followed by collapse to the homogeneous state. A narrow regime of oscillatory behaviour re-surfaces for even smaller $\Delta$, until finally, for $\Delta$ close to the up-switching point $\Delta_\uparrow$, the LDS instability manifests itself as the triggering of a wake of spatiotemporal chaos [region III]. Figure~\ref{Fig1}(d) shows simulated dynamics in this regime: the LDS proliferates into multiple copies that invade the surrounding homogeneous state. The dynamics can be interpreted as co-existence between a coherent metastable state (the lower branch homogeneous solution), and chaotic patterns ensuing from the Turing instability of the upper branch. The LDS triggers a (phase) transition from the former to the latter. We note that the resulting rise of a spatially localized (albeit expanding) domain of spatiotemporal chaos is different from the dynamics observed for driving frequencies $\Delta<\Delta_\uparrow$, where the lack of a stable (or metastable) homogeneous state forces the entire spatially extended field to simultaneously transform into de-localized spatiotemporal chaos. We also note that, although the discussion above involves a single driving strength $X=30$, the list of instabilities (periodic oscillations, chaotic oscillations, collapses, and transitions to spatiotemporal chaos) encompasses all dynamical behaviours identified for LDSs in the strong-driving regime of the AC-driven NLSE~\cite{leo_dynamics_2013}.

We have experimentally investigated LDS instabilities using a high-finesse optical fiber ring resonator that is coherently driven with an external laser. Prior studies in the weak-driving regime ($X<10$) have demonstrated that such a system is described by Eq.~\eqref{LLN}, and that it can support LDSs in the form of ``temporal cavity solitons'': persisting pulses of light that recirculate in the resonator~\cite{leo_temporal_2010,jang_ultraweak_2013,jang_temporal_2015,jang_controlled_2016}. In reference to Eq.~\eqref{LLN}, the ``spatial'' $x$ dimension represents a ``fast time'' variable ($x\rightarrow \tau$) that is defined in a reference frame moving with the LDSs, whilst $t$ is a ``slow time'' variable that describes the evolution of the optical field inside the resonator over several roundtrips. The dimensionless time-scales and the field amplitude in Eq.~\eqref{LLN} are related to the corresponding dimensional variables (indicated with subscript $D$) through the following normalisation~\cite{leo_temporal_2010,jang_controlled_2016}: $t = \alpha t_\mathrm{D}/t_\mathrm{R}$; $\tau = \tau_\mathrm{D} \sqrt{2\alpha/(|\beta_2|L)}$; $\psi = \psi_\mathrm{D}\sqrt{\gamma L/\alpha}$. Here, $t_\mathrm{R}$ is the cavity roundtrip time, $\alpha$ is equal to half of the percentage power dissipated over one round trip, $L$ is the length of the resonator, and $\beta_2 (<0)$ and $\gamma$ are the fiber group-velocity dispersion and nonlinearity coefficients, respectively. The normalised driving strength $S = (P_\mathrm{in}\gamma L\theta/\alpha^3)^{1/2}$, where $\theta$ is the intensity transmission coefficient of the coupler used to inject the coherent driving laser, with power $P_\mathrm{in}$, into the resonator. Finally, $\Delta$ characterizes the frequency detuning of the driving laser at $\omega$ from the closest cavity resonance at~$\omega_0$, with $\Delta \approx t_\mathrm{R} (\omega_0-\omega)/\alpha$.

\begin{figure}[t]
 \centering
  \includegraphics[width = \columnwidth, clip=true]{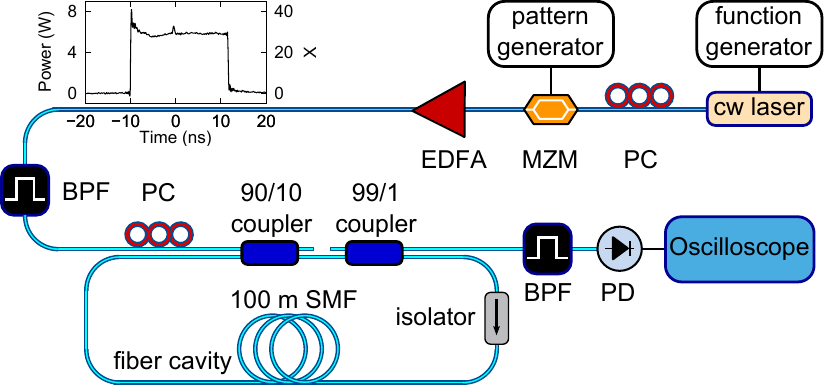}
 \caption{Schematic illustration of the experimental setup. PC: polarization controller; PD: photodetector; SMF: single-mode fiber. Top left inset shows a typical pump pulse profile.}
 \label{Fig2}
\end{figure}

Figure~\ref{Fig2} shows our specific experimental configuration. The fiber ring cavity consists of 100-meters of standard single-mode optical fiber with $\beta_2 = -21.4~\mathrm{ps^2km^{-1}}$ and $\gamma = 1.2~\mathrm{W^{-1}km^{-1}}$, closed on itself with a 90/10 fiber coupler. It also incorporates an optical isolator that inhibits stimulated Brillouin scattering, and a 99/1 tap-coupler through which the intracavity dynamics are monitored with a fast (12.5~GHz) photodetector and a real-time oscilloscope (12 GHz bandwidth). Before detection, the cavity output passes through a narrow (0.6~nm) bandpass filter (BPF) centered at 1551~nm. This removes the homogeneous background at 1550~nm, thereby improving the signal-to-noise ratio of the measurement~\cite{leo_temporal_2010}. The finesse of the cavity is $\mathcal{F} = \pi/\alpha\approx 20$.

Similar cavities have previously been employed to investigate stable LDSs~\cite{jang_ultraweak_2013, jang_temporal_2015, jang_controlled_2016}. In those studies, however, the cavity was driven with continuous wave (cw) laser light, which does not readily allow access to the strong driving regime required to study LDS instabilities. To overcome this issue, in our experiments we drive the cavity with approximately 20 ns flat-top pulses (see inset in Fig.~\ref{Fig2}) at a duty cycle of 25:1. Such quasi-cw pulses can easily be amplified to sufficiently high strengths, and we note that a similar pulsed pumping scheme was recently employed in a study of competing Faraday and Turing instabilities~\cite{copie_competing_2016}. We obtain our pump pulses by using a Mach-Zehnder modulator (MZM) to modulate a 1550~nm narrow linewidth cw laser. The MZM is driven with an electronic bit pattern generator at a frequency corresponding to an integer multiple of the cavity free-spectral range $t_\mathrm{R}^{-1}$, thus ensuring synchronous driving~\cite{coen_modulational_1997}. Before the pump pulses are coupled into the cavity through the 90/10 coupler ($\theta = 0.1$), they are amplified with an erbium-doped fiber amplifier (EDFA) to \emph{a quasi-cw} power of about $6~\mathrm{W}$ (normalised driving strength $X\sim 30$), and passed through a BPF centered at 1550~nm to remove amplified spontaneous emission.

\looseness=-1 To investigate different LDS regimes, we systematically control $\Delta$ by using a function generator to adjust the optical carrier frequency of the pump laser. Moreover, we monitor $\Delta$ by leveraging the non-ideal extinction of the MZM used to create the pump pulses: the linear resonance associated with the low-power cw background that exists in between the pump pulses provides for a $\Delta = 0$ reference. We estimate the error in $\Delta$ obtained in this way to be about $\pm 0.2$.  At this point, we emphasise that the 20~ns width of our quasi-cw pump pulses is about 4 orders of magnitude larger than the sub-picosecond widths of the LDSs that exist for our experimental parameters. The LDSs thus experience the driving as effectively homogeneous.

\begin{figure}[t]
 \centering
  \includegraphics[width = \columnwidth, clip=true]{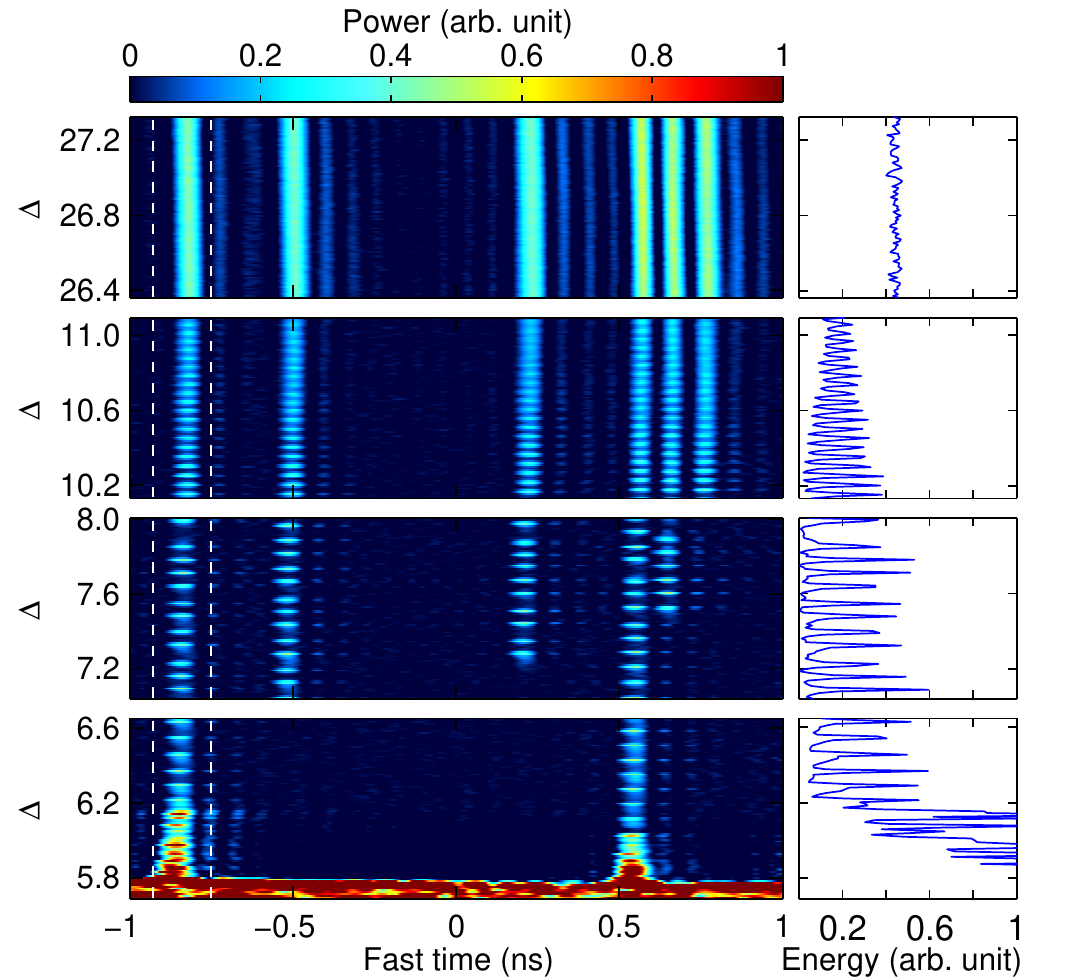}
 \caption{Experimentally measured spatiotemporal evolution inside the fiber ring resonator as $\Delta$ is continuously reduced. The different panels show snapshots around different $\Delta$ regions of interest, extracted from a single long measurement. Line plots on the right show the evolution of the integrated energy for a particular LDS (highlighted by dashed white lines).}
 \label{expr1}
\end{figure}

\looseness=-1Referring to Fig.~\ref{Fig1}, we start the experiment with the pump laser blue-detuned from a cavity resonance ($\Delta < 0$), and then slowly \emph{increase} $\Delta$ to $\Delta \approx 27$ by reducing the laser frequency. This results in the spontaneous excitation of LDSs~\cite{herr_temporal_2014, luo_spontaneous_2015}. We then reverse the direction of the frequency scan, and continuously \emph{reduce} $\Delta$ at a rate of about ${\mathrm{d}\Delta/\mathrm{d}t \approx -0.019~\mathrm{\mu s}^{-1}}$. (Similar forward + backward tuning has recently been used to control the number of stable LDSs in microresonators~\cite{karpov_universal_2016}.) Because the cavity photon lifetime ${t_\mathrm{ph}\approx1.6~\mathrm{\mu s}}$, the intracavity field reacts almost adiabatically to the frequency scan. Accordingly, by recording a long time trace at the cavity output, we are able to examine LDS behaviour as a function of $\Delta$. To facilitate visualisation, we divide the experimentally measured time trace into segments spanning one roundtrip, and concatenate the resulting sequences on top of each other. Figure~\ref{expr1} shows typical results obtained from such a measurement: the density maps depict the spatiotemporal field evolution (for clarity we only show a 2~ns segment of the full 20~ns pump profile), while the line plots capture the evolution of the integrated energy around a single LDS. The full measurement encompasses $\Delta$-values continuously reducing from about 27 to 5, but we only show snapshots around four regions that highlight the main dynamical behaviours.

The results are in remarkable agreement with the predicted bifurcation characteristics  [see Fig.~\ref{Fig1}(a)]. We first observe 6 stable LDSs until $\Delta$ passes the theoretically predicted Hopf threshold at about $\Delta\approx 11.3$, beyond which simple oscillatory behaviour ensues [region I in Fig.~\ref{Fig1}(a)]. For $\Delta < 8$, we witness irregular oscillations that lead to the spontaneous collapse of particular structures [region II]. Owing to the chaotic nature of the dynamics, and the fact that $\Delta$ is continuously reduced, some LDSs avoid a collapse and survive to the second regime of stable oscillations that manifests itself for $\Delta$ below the collapse region.  Finally, around $\Delta \approx 6$ the few remaining LDSs transform into localized chaotic domains [region III], which are quickly engulfed by the full destabilization of the coherent background at about $\Delta \approx 5.8$. This point coincides almost exactly with the theoretically predicted up-switching point $\Delta_\uparrow  = 5.75$.

Before proceeding, we comment on two aspects of our measurement. First, energy variations of oscillating LDSs are exaggerated in our experiment. This is due to the offset BPF in the detection path of our setup (see Fig.~\ref{Fig2}): oscillating LDSs exhibit variations in their spectral bandwidth, which can impact strongly on the energy transmitted through the BPF. Second, due to the 50~ps response time of our detection electronics, we are not able to directly discern the sub-picosecond profiles of the LDSs, or the proliferation into numerous closely-packed structures that are predicted in the chaotic regime [see Fig.~\ref{Fig1}(d)]. Nevertheless, the abrupt increase in integrated energy (and the spread of the LDS envelope) around $\Delta \approx 6$ provides convincing evidence for such behaviour.

\begin{figure}[htb]
 \centering
  \includegraphics[width = \columnwidth, clip=true]{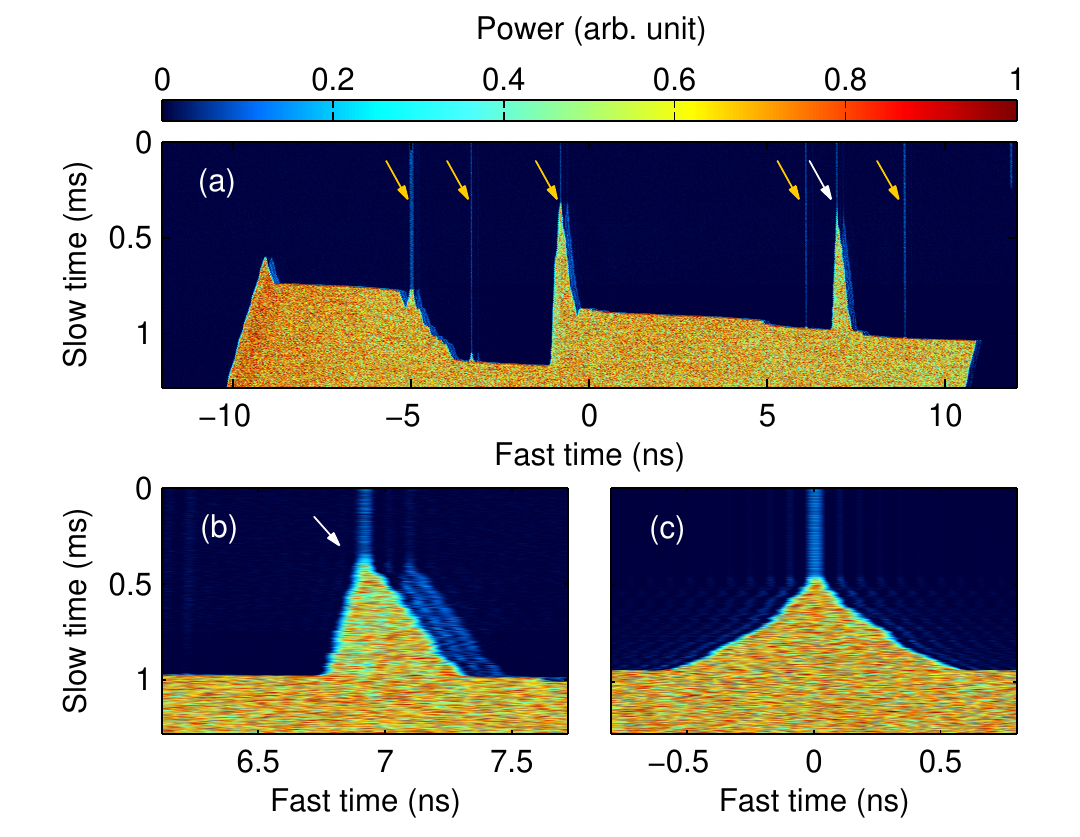}
 \caption{(a) Experimentally measured intracavity field evolution over the full 20~ns field profile as $\Delta$ is slowly reduced around $\Delta\approx 6$. Arrows highlight LDSs. (b) Zoom on a particular region (highlighted by a white arrow) where an LDS triggers an expanding domain of spatiotemporal chaos. Corresponding simulations are shown in (c).}
 \label{expr2}
\end{figure}

The results in Fig.~\ref{expr1} show clear evidence of oscillating and collapsing LDSs. However, due to the rate at which the driver frequency $\Delta$ was varied, the transitions to spatiotemporal chaos are not satisfactorily captured. We have therefore performed another experiment, where we significantly reduced the laser scan rate after reaching the region of interest ($\Delta\approx 6$). Figure~\ref{expr2}(a) shows the detected signal over the full 20~ns-width of the pump pulses when $\Delta$ is slowly reduced around $\Delta \approx 6$ over 1~ms (we estimate that $\Delta$ reduces from about 6.2 to 5.8 during the measurement). We observe 6 oscillating LDSs (highlighted with arrows), some of which abruptly transition into expanding fronts. Figure~\ref{expr2}(b) shows a zoom on a particular event [white arrow in Fig.~\ref{expr2}(a)], illustrating how the oscillating LDS triggers a chaotic domain that invades the surrounding homogeneous state at an almost constant rate. These observations are in good agreement with corresponding simulations of Eq.~\ref{LLN}, shown in Fig.~\ref{expr2}(c). The simulations use experimental parameters, aside from $\Delta$ which is reduced from 6.75 to 5.75 during the simulation (the values are nevertheless in reasonable agreement with experimental estimates). To facilitate comparison with experiments, the simulation results have been post-processed to mimic the effect of the offset filter and limited detection bandwidth in our experiment; the raw theoretical data is qualitatively similar to that shown in Fig.~\ref{Fig1}(d), consisting of chaotically oscillating, rapidly proliferating LDSs. In this context, besides results shown in Fig.~\ref{expr2}(c), we have carefully verified that all of our experimental observations are in good agreement with numerical simulations of both the AC-driven NLSE and a more rigorous Ikeda-like cavity map~\cite{coen_modeling_2013}.

The slow expansion of the domains triggered by individual LDSs is followed by an abrupt transition into fully distributed spatiotemporal chaos. This behaviour is explained by the driver frequency $\Delta$ falling below the up-switching point $\Delta_\uparrow\approx 5.8$ [see Fig.~\ref{Fig1}(a)]: the stable homogeneous state ceases to exist, thus forcing the entire field to switch to the turbulent Turing pattern on the upper branch. This transition takes place irrespective of the presence of LDSs. Indeed, we have performed additional experiments where no LDSs are present, and observed similar transition behaviour. We also note that the transition clearly highlights how inhomogeneities in the pump pulse strength (see inset in Fig.~\ref{Fig2}) can lead to locally different threshold behaviour. We suspect this explains why individual LDSs are observed to trigger the expanding chaotic domain at different $\Delta$. Besides pump inhomogeneities, the intrinsically chaotic oscillations of the LDSs can also contribute to the behaviour.

Our experimental results are in excellent agreement with the theoretically predicted LDS bifurcation characteristics. To the best of our knowledge, they represent the first experimental observations of LDSs exhibiting irregular oscillations, collapses and transitions into expanding domains of chaos. In this way, the findings directly confirm theoretical predictions that have been drawn across various physical contexts over the past three decades~\cite{barashenkov_existence_1996, nozaki_chaotic_1985, leo_dynamics_2013}. By implication, our results confirm that the AC-driven NLSE remains a valid predictor of Kerr nonlinear cavity dynamics in the strong-driving regime relevant to e.g. many microresonator frequency comb experiments~\cite{herr_temporal_2014, delhaye_optical_2007, grudinin_frequency_2012, webb_measurement_2016}. We accordingly predict that complex LDS instabilities may play a role in such systems. More generally, our work demonstrates that synchronously pumped fiber cavities permit systematic interrogation of complex LDS instabilities and their bifurcations. Given the universal nature of LDSs and related phenomenologies~\cite{cross_pattern_1993, coullet_localized_2002, akhmediev_dissipative_2005, akhmediev_dissipative_2008, purwins_dissipative_2010}, the ability to perform detailed and controlled laboratory experiments is expected to have wide impact across numerous nonlinear dissipative systems. For example, we note in closing that instabilities similar to those reported here have also been predicted for LDSs of the \emph{parametrically}-driven NLSE~\cite{bondila_topography_1995, alexeeva_taming_2001}, whose many applications include strongly coupled pendula~\cite{alexeeva_impurity-induced_2000}, vertically driven fluids~\cite{clerc_soliton_2009, wang_dynamics_1997}, nonlinear lattices~\cite{denardo_observations_1992}, and optical parametric oscillators~\cite{longhi_ultrashort-pulse_1995}.

\begin{acknowledgments}
We acknowledge support from the Marsden Fund and the Rutherford Discovery Fellowships administered by The Royal Society of New Zealand.
\end{acknowledgments}

\end{document}